\documentclass[11pt]{article}
\usepackage{amsmath,amssymb,amsbsy,amsfonts,amsthm,latexsym,
               amsopn,amstext,amsxtra,euscript,amscd,color}%,makecell}
\def \F {{\mathbb F}}
\def \Q {{\mathbb Q}}
\def \Z {{\mathbb Z}}

\def \V {{\mathbb V}}

\def \Tr {{\rm Tr_n}}
\def \T {{\rm Tr}}

\newtheorem{theorem}{Theorem}

\newtheorem{lemma}{Lemma}
\newtheorem{proposition}{Proposition}
\newtheorem{remark}{Remark}
\newtheorem{example}{Example}
\newtheorem{corollary}{Corollary}

\def\cB{{\mathcal B}}

\def\aa{{\bf a}}

\def\00{{\bf 0}}
\def\11{{\bf 1}}
\def\+{\oplus}

\def \F {{\mathbb F}}
\def \Q {{\mathbb Q}}
\def \Z {{\mathbb Z}}

\def \V {{\mathbb V}}

\def \Tr {{\rm Tr_n}}
\def \T {{\rm Tr}}

\begin{document}
\title{Full characterization of generalized bent functions as (semi)-bent spaces,
their dual, and the Gray image}
%and vectorial (affine) bent and semi-bent functions and spaces. \\
%A complete characterization! \\
%Gray images and duals and many other topics....}
%, their duals and Gray images for $q=2^k$}
\author{
Samir Hod\v zi\'c \footnote {University of Primorska, FAMNIT, Koper, Slovenia, e-mail: samir.hodzic@famnit.upr.si}\and
Wilfried Meidl \footnote{Johann Radon Institute for Computational and Applied Mathematics, OEAW, Linz, Austria,
email: meidlwilfried@gmail.com}\and
Enes Pasalic\footnote{University of Primorska, FAMNIT \& IAM, Koper, Slovenia, e-mail: enes.pasalic6@gmail.com}
}

\date{}
\maketitle

\begin{abstract}
In difference to many recent articles that deal with generalized bent (gbent) functions $f:\mathbb{Z}_2^n \rightarrow \mathbb{Z}_q$ for
certain small valued $q\in \{4,8,16 \}$, we give a complete description of these functions for both $n$ even and odd and for any $q=2^k$
in terms of both the necessary and sufficient conditions their component functions need to satisfy. This enables us to completely
characterize gbent functions as algebraic objects, namely as affine spaces of bent or semi-bent functions with interesting additional
properties, which we in detail describe.
%These conditions  are then equivalently given by identifying the gbent property so that the component functions  build  an affine space of
%bent functions with certain properties.
We also specify the dual and the Gray image of gbent functions
%regardless of the parity of $n$,
for $q=2^k$. We discuss the subclass of gbent functions which corresponds to relative difference sets, which we call $\Z_q$-bent functions,
and point out that they correspond to a class of vectorial bent functions.
%Their connection to relative difference sets is also addressed and for this purpose a notion of $2^k$-bent function is introduced.
The property of being $\Z_q$-bent is much stronger than the standard concept of a gbent function. We analyse
%but nevertheless we have been able to provide
two examples of this class of functions.

% This means that a function $f:\mathbb{Z}_2^n \rightarrow \mathbb{Z}_q$ represented as $f(x) = a_0(x) + 2a_1(x) + \cdots + 2^{k-1}a_{k-1}(x)$, where $q=2^k$ and
%$a_i:\mathbb{Z}_2^n \rightarrow \mathbb{Z}_2$, is gbent if and only if $\mathcal{A} = a_{k-1} \+ \langle a_0,a_1,\ldots,a_{k-2}\rangle$ is an affine vector space of
%bent functions such that for every (pairwise distinct) $g_i,g_j,g_l\in\mathcal{A}$
%the function $g_ig_j \+ g_ig_l \+ g_jg_l$ is bent.
%\bigskip
%\textbf{Keywords:}

\end{abstract}

\section{Introduction}\label{sec:pre}

Let $\V_n$ be an $n$-dimensional vector space over $\F_2$, and let $\mathcal{B}_n$ denote the set of Boolean functions from $\V_n$ to $\F_2$.
The {\it Walsh-Hadamard transform} of $f\in\mathcal{B}_n$ at a point $u  \in\V_n$ is defined by
$$\mathcal{W}_{f}(u)=\sum_{x\in \V_n}(-1)^{f(x)\oplus  u\cdot x},$$
where ``$\cdot$'' denotes an inner product on $\V_n$. When $\V_n = \F_2^n$ we may take the dot product,
%The classical frameworks are $\V_n=\F_2^n$, in which case we can choose
%the conventional dot product as inner product \textcolor[rgb]{0.98,0.00,0.00}{(i.e, $\langle u,x\rangle=x \cdot y$, $x,y\in \V_n$)}, and $\V_n=\F_{2^n}$,
in the case of a finite field with $2^n$ elements $\F_{2^n}$, the standard inner product is
%correspond to (using a suitable basis)
$ x \cdot y = \Tr(xy)$, where $\Tr(z)$ denotes the absolute trace of $z$.
Note that the characters of $\V_n\times\F_2$ are
$\chi_{a,u}(x,y) = (-1)^{ay \+  u \cdot x}$, $a\in\{0,1\}$, $u\in\V_n$. Hence $\mathcal{W}_{f}(u) = \chi_{1,u}(D)$, where
$D=\{(x,f(x))\,:\,x\in\V_n\}$ is the graph of $f$.

A function $f\in\mathcal{B}_n$ is called {\it bent} if $\mathcal{W}_{f}(u)$ has absolute value $2^{n/2}$ for all $u\in\V_n$, i.e.,
$|\chi_{a,u}(D)| = 2^{n/2}$ for all characters which are nontrivial on $\{0\}\times\F_2$ ($a\ne 0$), hence $f$ can be identified with a
relative difference set $D$ in $\V_n\times\F_2$, see \cite{p,tpf}. Since $\mathcal{W}_{f}(u)$ is an integer, for a bent function we have
$\mathcal{W}_{f}(u) = 2^{n/2}(-1)^{f^*(u)}$ for a Boolean function $f^*\in\mathcal{B}$, called the {\it dual} of $f$, which then is also bent.
Obviously, Boolean bent functions only exist when $n$ is even. When $n$ is odd, a {\it semi-bent} function is defined as a function $f\in\mathcal{B}_n$
for which $\mathcal{W}_f(u) \in\{\pm 2^{\frac{n+1}{2}}, 0\}$ for all $u\in\V_n$. A function $f\in \mathcal{B}_n$ is called \emph{$s$-plateaued} if its
Walsh spectrum only takes three values $0$ and $\pm 2^{\frac{n+s}{2}}$ ($0\le s\le n$). Note that $n$ and $s$ must have the same parity.
%\textcolor[rgb]{0.98,0.00,0.00}{(I think the usual definition is this.)}
Let $K$ be a subset of $\V_n$. By
$\phi_{K}$ we denote a Boolean function in $\mathcal{B}_n$ whose value in $x$ is $1$ if $x\in K$ and $0$ if $x\not\in K$. The function $\phi_E$ is called
\emph{the indicator of $K$}.

Many more variants of bent functions, like bent functions in odd characteristic, vectorial bent functions from $\F_p^n$ to $\F_p^m$, negabent
functions, bent$_4$ functions, all corresponding to relative difference sets in respective groups, have been investigated. The reader is referred to, for instance,   the
articles \cite{gps,ksw,pp,psz,sz,z} and the recent survey article \cite{p16}. For a very general viewpoint considering bent functions over arbitrary
abelian groups, we refer to \cite{p04}.

For a positive integer $q$ let $\Z_q$ be the ring of integers modulo $q$.
We call a function $f$ from $\V_n$ to $\Z_q$ a {\it generalized Boolean function}, and denote the set of generalized Boolean functions
from $\V_n$ to $\Z_q$ by $\mathcal{GB}_n^q$. If $q=2$, then $f$ is Boolean, and $\mathcal{GB}_n^2 = \mathcal{B}_n$.
%The set of all Boolean functions in $n$ variables, that is the mappings from $\mathbb{Z}_2^n$ to $\mathbb{Z}_2$ is denoted by $\mathcal{B}_n$.
Having applications of functions from $\V_n$ to $\Z_4$ in code-division multiple access systems in mind, in \cite{kus} Schmidt introduced a class
of functions which further on were called {\it generalized bent} ({\it gbent}). A function $f\in \mathcal{GB}^q_n$ for which the {\em generalized
Walsh-Hadamard transform} (GWHT) at a point $u\in\V_n$ defined as the complex valued function
$$\mathcal{H}^{(q)}_{f}(u)=\sum_{x\in \V_n}\zeta_q^{f(x)}(-1)^{ u\cdot x},$$
where $\zeta_q=e^{2\pi i/q}$ (or any other complex $q$th-primitive root of unity), has absolute value $2^{n/2}$ for all $u\in\V_n$, is called a
%(sometimes we denote $\zeta_q=\zeta$)
generalized bent function. Note that when $f$ is Boolean, then $\mathcal{H}^{(2)}_{f}(u) = \mathcal{W}_f(u)$.

Currently there is a lot of research activity regarding the construction and analysis of gbent functions, see for instance
\cite{SH2,lff,Gray,Vect,m,kus,Tok,smgs}. The quaternary $q=4$ and octal case $q=8$ were investigated in \cite{kus,Tok} and \cite{m,Octal}, respectively.
Quite recently \cite{Gray}, a complete characterization of gbent functions was also given for $q=16$. In these cases gbent functions were fully characterized
by specifying both the related necessary and sufficient conditions. The first general characterization of gbent functions, in terms of the coordinate
functions for any $q$ even and $n$ both even/odd, is given in \cite[Theorem 4.1]{SH2}. More precisely, for $2^{k-1}<q\leq 2^k,$ to any generalized function
$f:\mathbb{Z}^n_2\rightarrow \mathbb{Z}_q,$ one may associate a unique sequence of Boolean functions $a_i\in \mathcal{B}_n$ ($i=0,1,\ldots,h-1$) such that
$f(x)=a_0(x)+2a_1(x)+2^2a_2(x)+\ldots+2^{k-1}a_{k-1}(x),\; \forall x\in \mathbb{Z}^n_{2}$, where $a_i$ are then called the coordinate functions of $f$.
The gbent conditions derived in \cite{SH2} were only sufficient and whether these are also necessary was left as an open problem.
Necessary conditions for a function $f\in\mathcal{GB}_n^{2^k}$ to be gbent are in \cite{Gray} when $n$ is even and also for odd $n$ in \cite{mmms}.

In this paper we are interested in gbent functions in $\mathcal{GB}_n^{q}$ where $q=2^k$ and $k>1$ is a positive integer. In difference to the above
mentioned results that only considered gbent functions for $q \in \{4,8,16\}$, we give a complete characterization of these functions for both $n$
even and odd and for any $q=2^k$. We start with with both, {\it necessary and sufficient} conditions their coordinate functions need to satisfy.
These conditions are equivalent to those very recently published online in \cite{txqf}.
Notably we then describe gbent functions as algebraic objects, a characterization which goes far beyond the conventional descriptions in terms the
Walsh transforms of linear combinations of the coordinate functions, which in accordance with the terminilogy for vectorial bent function we
call the {\it component functions} of the gbent function. We show that gbent functions correspond to affine spaces of bent functions
when $n$ is even and semi-bent functions when $n$ is odd, with certain interesting additional properties, which we precisely describe.
Employing conventional equivalence, we show that gbent functions and affine spaces of bent (semi-bent) functions with these properties,
are identical objects. These results essentially completely resolve the case of gbent functions from $\V_n$ to $\Z_{2^k}$, using the approach
based on Hadamard matrices introduced in \cite{SH2}.

We recall that in the case of $q=2^k$ we always have $\mathcal{H}_f^{(2^k)}(u) = 2^{n/2}\zeta_{2^k}^{f^*(u)}$,
(except for the case that $n$ is odd and $q=4$), for a function $f^*\in\mathcal{GB}_n^{2^k}$, which we call the dual of $f$, see \cite{Gray}. As
pointed out in \cite{Vect}, $f^*$ is also a gbent function. In this direction, we completely specify the dual and Grey image of any gbent function when
$n$ is even and for any $q=2^k$. The case $n$ being odd appears to be harder to deal with the approach based on Hadamard matrices (as used in this article)
and it is left as an open problem. Regarding a natural connection of gbent functions and the so-called Gray images, it is shown that the Gray map
%$\varphi(f)\in\mathcal{B}_{n+k-1}$
of a gbent function $f \in\mathcal{GB}_n^{2^k}$ is $(k-1)$-plateaued if $n$ is even, and $(k-2)$-plateaued if $n$ is odd. This generalizes
the results on the Gray map given in \cite{Gray,Tok} for $k=2,3$ and $4$.

%For some results on functions from $\V_n$ to $\Z_q$ for general $q$ we refer to \cite{lff,SH2}.
%
%We recall that in the case of $q=2^k$ we always have $\mathcal{H}_f^{2^k}(u) = 2^{n/2}\zeta_{2^k}^{f^*(u)}$,
%(except for the case that $n$ is odd and $q=4$), for a function $f^*\in\mathcal{GB}_n^{2^k}$, which we again
%call the dual of $f$, see \cite{Gray}. As pointed out in \cite{Vect}, $f^*$ is also gbent. Differently to the case $k = 1$, gbent functions
%also exist for odd $n$ when $k > 1$.

We emphasize here that a gbent function conceptually {\it does not} correspond to a bent function, since in the definition of GWHT not all
characters of $\V_n\times\Z_{2^k}$ are considered. Thus, in general, a gbent function does not give rise to
a relative difference set. For this reason we extend the definition and introduce the term of a $\Z_q$-bent function.
We call a function $f\in\mathcal{GB}_n^{2^k}$ a {\it $\Z_q$-bent} function if
$$\mathcal{H}^{(q)}_{f}(a,u)=\sum_{x\in \V_n}\zeta_q^{af(x)}(-1)^{ u\cdot x}$$
has absolute value $2^{n/2}$ for all $u\in\V_n$ and all nonzero $a\in\Z_{2^k}$.
Whereas there are several constructions of gbent functions, a class of $\Z_q$-bent functions seems not to be easy to obtain.
For a construction using (partial) spreads, we refer to \cite{Vect}. \\[.3em]
%
%{\bf Nevertheless, we provide two examples of this class of functions though generic construction methods still need to be specified. As a consequence, we also %obtain relative difference
%sets related to the structure $\V_n\times\Z_{2^k}$. It is not clear whether we introduce the term $2^k$-bent function or it was already known ? We mention the %construction in \cite{Vect},
%so these functions were defined there  ??}
%\textcolor[rgb]{0.98,0.00,0.00}{It is right, this type of gbent functions is considered in \cite{Vect} - but called a ``gbent function which is vectorial'' - in %connection with
%difference sets. The definition is not new anyway, it only is a special case of the general definition of a bent function between 2 groups with their character %groups.}
%

This article is organized as follows. In Section \ref{prel} we recall some preliminary results which are used later.
A necessary and sufficient condition, given in terms of Hadamard matrices, for a function $f\in\mathcal{GB}_n^{2^k}$ to be gbent is given in Section \ref{ifif}.
%This continues the research in \cite{SH2} where Hadamard matrices were used for the first time to analyse gbent
%functions.
In Section \ref{chara}, we use these conditions to completely characterize gbent functions as affine (semi-)bent spaces
with certain properties.
%For $n$ even ($n$ odd), a gbent function is an affine space of bent functions (semibent functions)
%with certain interesting additional properties, and vice versa. We precisely describe these additional properties.
We introduce the term $\Z_q$-bent functions for those gbent functions that correspond to relative difference sets in
$\V_n\times\Z_q$, $q=2^k$, in Section \ref{RDS}. We describe some of their properties and analyse two
%To relate gbent functions to relative difference sets the concept of $2^k$ bent functions is introduced in Section \ref{RDS} and their
%connection to gbent functions is specified. Two
explicit examples of this class of functions.
%are also provided here.
%as we point out in Section \ref{RDS}, As a consequence of this result, a $2^k$-bent function $f \in\mathcal{GB}_n^{2^k}$ is a certain
%vectorial bent function of dimension $k$.
In Section \ref{dual} we specify the dual and Gray map of gbent functions $f\in\mathcal{GB}_n^{2^k}$.

\section{Preliminaries}
\label{prel}

A $(1,-1)$-matrix $H$ of order $p$ is called a \emph{Hadamard} matrix if  $HH^{T}=pI_p,$ where $H^{T}$ is the transpose of $H$, and $I_p$ is the $p\times p$ identity matrix. A special kind of  Hadamard matrix is
the \emph{Sylvester-Hadamard} or \emph{Walsh-Hadamard} matrix, denoted by $H_{2^{k}},$ which is constructed recursively using Kronecker product $H_{2^{k}}=H_2\otimes H_{2^{k-1}},$ where
\begin{eqnarray*}\label{HM}
H_1=(1);\hskip 0.4cm H_2=\left(
                           \begin{array}{cc}
                             1 & 1 \\
                             1 & -1 \\
                           \end{array}
                         \right);\hskip 0.4cm H_{2^k}=\left(
      \begin{array}{cc}
        H_{2^{k-1}} & H_{2^{k-1}} \\
        H_{2^{k-1}} & -H_{2^{k-1}} \\
      \end{array}
    \right).
\end{eqnarray*}
For technical reasons we start the row and column index of $H_{2^k}$ with $0$, and we denote the $r$-th row of $H_{2^k}$ by
$H_{2^k}^{(r)}$, $0\le r\le 2^k-1$. To an integer $j=\sum_{i=0}^{k-1}j_i2^i$, $0\le j\le 2^k-1$, we assign
$z_j = (j_0,j_1,\ldots,j_{k-1})\in\F_2^k$, which also implies an ordering of the elements of $\F_2^k$.

We summarize some properties of the Sylvester-Hadamard matrix in the following lemma. The first one follows from
the recursive definition of $H_{2^k}$, the second is the well-known property that each row of $H_{2^k}$ is the
evaluation of some linear function. The third one may be less well known, hence we provide the proof of this property.
\begin{lemma}
\label{Hada}
\begin{itemize}
\item[(i)] Each row of $H_{2^k}$ is uniquely determined by the signs of the entries at  positions $2^s$, $s=0,1,\ldots,k-1$.
\item[(ii)] Let $z_j = (j_0,j_1,\ldots,j_{k-1})\in\F_2^k$, where $j=\sum_{i=0}^{k-1}j_i2^i$, $0\le j\le 2^k-1$. Then
\[ H_{2^k}^{(r)} = ((-1)^{z_0\cdot z_r}, (-1)^{z_1\cdot z_r}, \ldots, (-1)^{z_{2^k-1}\cdot z_r}). \]
\item[(iii)] Let $W = (w_0,w_1,\ldots,w_{2^k-1})$, where $w_i = \pm 1$, $0\le i\le 2^k-1$. Then $W = \pm H_{2^k}^{(r)}$
for some $r\in\{0,\ldots,2^k-1\}$ if and only if for any four distinct integers $j,c,l,v \in \{0,\ldots,2^k-1\}$
such that $z_j \+ z_c\+ z_l\+ z_v = \00$ we have
\begin{equation}
\label{wwww}
w_jw_c = w_lw_v.
\end{equation}
\end{itemize}
\end{lemma}
{\it Proof of (iii).}
Let $W = (w_0,w_1,\ldots,w_{2^k-1}) = \pm H_{2^k}^{(r)}$ for some (fixed) $r\in\{0,\ldots,2^k-1\}$, and let
$j,c,l,v\in\{0,\ldots,2^k-1\}$ be arbitrary distinct integers such that $z_j\oplus z_c\oplus z_l\oplus z_v={\bf 0}$.
By (ii),
\[ H^{(r)}_{2^k}=((-1)^{z_r\cdot z_0},(-1)^{z_r\cdot z_1},\ldots,(-1)^{z_r}\cdot z_{2^k-1}). \]
Hence relation $(\ref{wwww})$ can be written as
%Since we have $\mathcal{W}(u)=(W_0(u),\ldots,W_{2^p-1}(u))=\pm 2^{\frac{n}{2}} H^{(r)}_{2^{k-1}}$, neglecting the signs "$\pm$" (what does not reduce the generality), the relation (\ref{wquad}) divided by $2^{\frac{n}{2}}$ we may write as
\[ (-1)^{z_r\cdot z_j}(-1)^{z_r\cdot z_c}=(-1)^{z_r\cdot z_{l}}(-1)^{z_r\cdot z_{v}}, \]
or equivalently
\[ (-1)^{z_r\cdot (z_j\oplus z_c \oplus z_l \+ z_v)} = 1, \]
which is satisfied for $z_j,z_c,z_l,z_v$ with $z_j\oplus z_c\oplus z_l\oplus z_v={\bf 0}$. \\
Suppose conversely that $(\ref{wwww})$ holds for all $j,c,l,z,v$ with $z_j\oplus z_c\oplus z_l\oplus z_v=\00$.
Then, to show that  $W = \pm H_{2^k}^{(r)}$ for some $r$, $0\le r\le 2^k-1$, we proceed by induction on $k$.
Trivially it holds for $k=1$, since $\pm H_2^{(r)}$, $r=0,1$, covers all possible combinations for $(w_0,w_1)$.
For $k=2$, we first notice that all solutions of the equality $w_0w_1=w_2w_3$ with $w_i = \pm 1, i=0,1,2,3$, are the
quadruples $(w_0,w_1,w_2,w_3)$ containing an even number of $-1$s. As it is easy to see, all such quadruples $W$
are of the form $W = (w_0,w_1,\pm(w_0,w_1))$, hence equal to $\pm H_4^{(r)}$ for some $r\in \{0,1,2,3\}$.
Before we continue with the induction proof, we also add the argument for $k=3$. With the above argument applied to
the quadruples $(4,5,6,7)$ and $(0,1,4,5)$, we get $(w_6,w_7) = \pm (w_4,w_5)$ and $(w_4,w_5) = \pm (w_0,w_1)$.
Consequently,
\begin{eqnarray}\label{sim} \nonumber
(w_0,\ldots,w_7) &=& (w_0,w_1,\pm (w_0,w_1),\pm (w_0,w_1,\pm(w_0,w_1))) \\ \nonumber
%&=&(\mathcal{W}_{g_{0}},\ldots,\mathcal{W}_{g_{3}},\pm (\mathcal{W}_{g_{0}},\ldots,\mathcal{W}_{g_{3}}))\\
&=&\pm (H^{(d)}_{2^2},\pm H^{(d)}_{2^2})=\pm H^{(r)}_{2^3},
\end{eqnarray}
for some $0 \le r\le 7$.  \\
Now suppose that the following holds for a tuple $W = (w_0,w_1,\ldots,w_{2^{k-1}-1})$ of length $2^{k-1}$ with entries in $\{-1,1\}$:
If for all $0\le j < c < l < v\le 2^{k-1}-1$ with $z_j \+ z_c \+ z_l \+ z_v = {\bf 0}$ we have $w_jw_c = w_lw_v$, then
$W = \pm H_{2^{k-1}}^{(r)}$ for some $r \in\{0,1,\ldots,2^{k-1}-1\}$.

Let now $W = (w_0,w_1,\ldots,w_{2^k-1})$, $w_i = \pm 1$, $i=0,1,\ldots,2^k-1$, such that $w_jw_c = w_lw_v$ for all
$0\le j < c < l < v\le 2^k-1$ with $z_j \+ z_c \+ z_l \+ z_v = {\bf 0}$. By induction hypothesis, we then have
$(w_0,w_1,\ldots,w_{2^{k-1}-1}) = \pm H_{2^{k-1}}^{(r)}$ and $(w_{2^{k-1}},w_{2^{k-1}+1},\ldots,w_{2^k-1}) = \pm H_{2^{k-1}}^{(\bar{r})}$
for some $r,\bar{r} \in \{0,1,\ldots,2^{k-1}-1\}$. We have to show that $\bar{r} = r$, or equivalently
$w_{2^{k-1}+j} = w_j$, $j=0,1,\ldots, 2^{k-1}-1$, or $w_{2^{k-1}+j} = -w_j$, $j=0,1,\ldots, 2^{k-1}-1$. By (i), it is sufficient to show
that $w_{2^{k-1}+2^s} = w_{2^s}$, $s=0,1,\ldots,k-2$, or $w_{2^{k-1}+2^s} = -w_{2^s}$, $s=0,1,\ldots,k-2$.
We consider the quadruples $(j,c,l,v) = (0,2^s,2^{k-1},2^{k-1}+2^s)$, $s=0,1,\ldots,k-2$, for which $z_j \+ z_c \+ z_l \+ z_v = 0$ always
holds. Since they satisfy $(\ref{wwww})$, either
$w_0,w_{2^s},w_{2^{k-1}},w_{2^{k-1}+2^s}$ have all the same sign, or exactly two of them are negative. Consequently, if
$w_0 = w_{2^{k-1}}$,
then we must have $w_{2^s} = w_{2^{k-1}+2^s}$, $s=0,1,\ldots,k-2$, and if $w_0 = -w_{2^{k-1}}$, then $w_{2^s} = -w_{2^{k-1}+2^s}$, $s=0,1,\ldots,k-2$.
\hfill$\Box$ \\[.5em]
In what follows we  derive and recall some basic results on gbent functions, which are proved useful  in the sequel.
%and summarize some of the recent achievements on the analysis of gbent functions.
\begin{lemma}
\label{sqrt2}
Let $k\ge 3$. Then $\sqrt{2}\zeta_{2^k}^j$ is uniquely represented in $\Q(\zeta_{2^k})$ as
\[ \sqrt{2}\zeta_{2^k}^j = \pm \zeta_{2^k}^{J_1} \pm \zeta_{2^k}^{J_2} \in \Q(\zeta_{2^k}). \]
for some $0\le J_1 < J_2 \le 2^{k-1}-1$ with $J_2-J_1 = 2^{k-2}$.
\end{lemma}
{\it Proof.} W.l.o.g. let $\zeta_{2^3} = \zeta_{2^k}^{2^{k-3}} = (1+i)/\sqrt{2}$, and hence
\[ \sqrt{2}\zeta_{2^k}^j = (\zeta_{2^k}^j + i\zeta_{2^k}^j)/\zeta_{2^k}^{2^{k-3}} = \zeta_{2^k}^{j-2^{k-3}} + \zeta_{2^k}^{2^{k-2}}\zeta_{2^k}^{j-2^{k-3}} =
\zeta_{2^k}^{j-2^{k-3}} + \zeta_{2^k}^{j+2^{k-3}}. \]
As $\zeta_{2^k}^j = -\zeta_{2^k}^{j-2^{k-1}}$ we can assume that $0\le j\le 2^{k-1}-1$. Again using that $\zeta_{2^k}^{2^{k-1}} = -1$, we can then write
$\sqrt{2}\zeta_{2^k}^j$ as
\[ \sqrt{2}\zeta_{2^k}^j =
\left\{
\begin{array}{ll}
-\zeta_{2^k}^{j-2^{k-3}+2^{k-1}} + \zeta_{2^k}^{j+2^{k-3}} & \mbox{if}\; j-2^{k-3}<0, \\
\zeta_{2^k}^{j-2^{k-3}} + \zeta_{2^k}^{j+2^{k-3}} & \mbox{if}\; 0\le j-2^{k-3} < j+2^{k-3} < 2^{k-1}, \\
\zeta_{2^k}^{j-2^{k-3}} - \zeta_{2^k}^{j+2^{k-3}-2^{k-1}} & \mbox{if}\; j+2^{k-3} \ge 2^{k-1}.
%-\zeta_{2^k}^{j-2^{k-3}-2^{k-1}} - \zeta_{2^k}^{j+2^{k-3}-2^{k-1}} & \mbox{if}\; 0\le j-2^{k-3} > 2^{k-1}.
\end{array}
\right.
\]
In either case $\sqrt{2}\zeta^j$ is of the form $\pm \zeta_{2^k}^{J_1} \pm \zeta_{2^k}^{J_2}$ for some $0\le J_1 < J_2 \le 2^{k-1}-1$ with $J_2-J_1 = 2^{k-2}$.
Since $\{1,\zeta_{2^k},\ldots,\zeta_{2^k}^{2^{k-1}-1}\}$ is a basis of $\Q(\zeta_{2^k})$, this representation is unique.
\hfill$\Box$\\[.5em]
To any generalized Boolean function $f:\V_n\rightarrow \Z_{2^k},$ we may associate the sequence of Boolean functions $a_j\in \mathcal{B}_n$, $j=0,1,\ldots,k-1$, for which
\begin{eqnarray}\label{eq:1}
f(x)=a_0(x)+2a_1(x)+2^2a_2(x)+\cdots+2^{k-1}a_{k-1}(x),\; \forall x\in \V_n.
\end{eqnarray}
% Let $f\in\mathcal{GB}_n^{2^k}$ be given as
% \[ f(x) = a_0(x)+2a_1(x)+\cdots+2^{k-2}a_{k-2}(x)+2^{k-1}a_{k-1}(x), \quad a_j\in\mathcal{B}_n,\, 0\le j\le k-1. \]
For an integer $i$, $0\le i\le 2^{k-1}-1$, with $i = \sum_{j=0}^{2^{k-1}}i_j2^j$, $i_j\in\{0,1\}$, we define the
$i$-th {\it component function} $g_i\in\mathcal{B}_n$ of $f$ as
\begin{equation}
\label{comfu}
g_i(x) = a_{k-1}(x)\+i_0a_0(x)\+\cdots \+i_{k-2}a_{k-2}(x).
\end{equation}
For an element $u\in\V_n$, let $\mathbf{\mathcal{W}}(u) = (\mathcal{W}_{g_0}(u),$ $\mathcal{W}_{g_1}(u),\ldots,\mathcal{W}_{g_{2^{k-1}-1}}(u))$ and
let $\mathbf{S}(u) = (S_0,S_1,\ldots,S_{2^{k-1}-1})$ be the vector defined by
\begin{eqnarray}\label{Sk2}
\mathbf{S}(u)=\left(
    \begin{array}{c}
      S_0 \\
      S_1 \\
      \vdots \\
      S_{2^{k-1}-1} \\
    \end{array}
  \right)
:=
H_{2^{k-1}}\left(\begin{array}{c}
                            \mathcal{W}_{g_0}(u) \\
                           \mathcal{W}_{g_1}(u) \\
                           \vdots \\
                           \mathcal{W}_{g_{2^{k-1}-1}}(u) \\
                           \end{array}
                         \right).
\end{eqnarray}

In \cite{SH2} the following proposition has been shown.
%it has been shown \cite{SH2} that the GWHT of the function $f$ at point $u\in \mathbb{Z}^n_2$ can be written as $\mathcal{H}_f(u)=2^{-p}(S^T B),$ where $B=[\zeta^k]^{2^{p}-1}_{k=0}$ and $S^T$
%is the transpose of $S$. Note that both matrices $S$ and $W$ depend on the input $u\in \mathbb{Z}^n_2,$ since they contain WHTs $W_i(u),$ $i=0,\ldots,2^{p}-1.$ We recall the following result.
\begin{proposition}\cite{SH2}
\label{HfromS}
Let $f\in\mathcal{GB}_n^{2^k}$ and $u\in\V_n$. Then
\[ 2^{k-1}\mathcal{H}_f(u) = (1,\zeta_{2^k},\ldots,\zeta_{2^k}^{2^{k-1}-1})\cdot \mathbf{S}(u) = S_0+S_1\zeta_{2^k}+\cdots+S_{2^{k-1}-1}\zeta_{2^k}^{2^{k-1}-1}. \]
\end{proposition}

\section{Necessary and sufficient conditions}
\label{ifif}

In this section we present necessary and sufficient conditions for the gbentness of functions $f\in\mathcal{GB}_n^{2^k}$ given as in $(\ref{eq:1})$.
We provide an  equivalent form of these conditions in terms of certain spectral properties of the component functions of $f$. In the next section, we will use
these conditions to completely characterize gbent functions as algebraic objects, which are shown to possess a lot of structure and to have some interesting
properties.
%
% In this section we first give a complete characterization of gbent functions for $q=2^k$ in terms of necessary and sufficient conditions.
% Furthermore, we analyze gbent conditions involved in this characterization in terms of duals of linear combinations of component functions,
% where we provide several equivalent forms. First, we start with the characterization.
\begin{theorem}
\label{iff1}
Let $f(x) = a_0(x)+\cdots+2^{k-2}a_{k-2}(x)+2^{k-1}a_{k-1}(x) \in\mathcal{GB}_n^{2^k}$, and let
$g_i(x) = a_{k-1}(x) \+ i_0a_0(x)\+i_1a_1(x)\+\cdots\+i_{k-2}a_{k-2}(x)$, $0\le i\le 2^{k-1}-1$, where
$i=\sum_{j=0}^{k-2}i_j2^j$ and $i_j \in \{0,1\}$.
\begin{itemize}
\item[(i)]
If $n$ is even, then $f$ is gbent if and only if $g_i$ is bent for all $0\le i\le 2^{k-1}-1$, such that
for all $u\in\V_n$,
\begin{equation}\label{W=S}
\mathbf{\mathcal{W}}(u) = (\mathcal{W}_{g_0}(u),\mathcal{W}_{g_1}(u),\ldots,\mathcal{W}_{g_{2^{k-1}-1}}(u)) = \pm 2^{\frac{n}{2}}H_{2^{k-1}}^{(r)}
\end{equation}
for some $r$, $0\le r\le 2^{k-1}-1$, depending on $u$.
\item[(ii)]
If $n$ is odd, then $f$ is gbent if and only if $g_i$ is semi-bent for all $0\le i\le 2^{k-1}-1$, such that
for all $u\in\V_n$,
\begin{equation}
\label{oddW=S}
\mathbf{\mathcal{W}}(u) = (\pm 2^{\frac{n+1}{2}}H^{(r)}_{2^{k-2}},\textbf{0}_{2^{k-2}})\quad\mbox{or}\quad
\mathbf{\mathcal{W}}(u) = (\textbf{0}_{2^{k-2}},\pm 2^{\frac{n+1}{2}} H^{(r)}_{2^{k-2}})
\end{equation}
for some $r$, $0\le r\le 2^{k-2}-1$, depending on $u$ ($\textbf{0}_{2^{k-2}}$ is the all-zero vector of length $2^{k-2}$).
\end{itemize}
\end{theorem}
\begin{proof} First we consider the case (i) when $n$ is even.
The sufficiency of $(\ref{W=S})$ has been shown in \cite{SH2}, though in a more general context for $f\in\mathcal{GB}_n^q$, where $q$ is an arbitrary even
integer. For the sake of completeness we include the proof arguments here.
Suppose that  $(\ref{W=S})$ holds, which also implies that all $g_i$ are bent. By the definition of $S_t$, $0\le t\le 2^{k-1}-1$, we then
have $S_t = 0$ if $t\ne r$, and $S_r = \pm 2^{n/2}2^{k-1}$. Proposition \ref{HfromS} then yields
$2^{k-1}\mathcal{H}_h(u) = \pm 2^{n/2}2^{k-1}\zeta_{2^k}^r$, hence $f$ is gbent.

Now, conversely,  suppose that $f$ is gbent. By Proposition \ref{HfromS}, we then have
%By Proposition \ref{mmsres} all compoenets $g_i$, $0\le i\le 2^{k-1}-1$, are then bent, i.e.,
%$\mathbf{\mathcal{W}}(u) = 2^{n/2}(\pm 1,\pm 1,\ldots,\pm 1)$. Furthermore,
\[ S_0 + S_1\zeta_{2^k} + \cdots + S_{2^{k-1}-1}\zeta_{2^k}^{2^{k-1}-1} = 2^{k-1}\mathcal{H}_f(u) = \pm 2^{k-1}2^{\frac{n}{2}}\zeta_{2^k}^r \]
for some $r$, $0\le r\le 2^{k-1}-1$. Since $\{1,\zeta_{2^k},\ldots,\zeta_{2^k}^{2^{k-1}-1}\}$ is a basis of $\Q(\zeta_{2^k})$, this implies that $S_t = 0$, $0\le t\le 2^{k-1}-1$,
$t \ne r$, and $S_r = \pm 2^{k-1}2^{\frac{n}{2}}$. By the invertibility of $H_{2^{k-1}}$, the only solution for $\mathbf{\mathcal{W}}(u)$ in the resulting linear
system is $\mathbf{\mathcal{W}}(u) = \pm 2^{\frac{n}{2}}H_{2^{k-1}}^{(r)}$. Hence $(\ref{W=S})$ holds, also implying that all $g_i$ are bent.

For the case (ii), when $n$ is odd, the sufficiency of  $(\ref{oddW=S})$  has also been shown in \cite{SH2}. Again, for the sake of completeness, we
include the proof arguments. If $(\ref{oddW=S})$ holds, then by $(\ref{Sk2})$, for $j \in \{ r,r+2^{k-2}\}$ we have $S_j = \pm 2^{k-2}2^{\frac{n+1}{2}}$, and
$S_j = 0$ if $j\ne r,r+2^{k-2}$. Hence,  from Proposition \ref{HfromS}, we get
\[ \mathcal{H}_f(u) = \pm 2^{\frac{n+1}{2}}\zeta_{2^k}^r \pm 2^{\frac{n+1}{2}}\zeta_{2^k}^{r+2^{k-2}} = 2^{\frac{n+1}{2}}\zeta_{2^k}^r(\pm 1 \pm i) = 2^{\frac{n}{2}}\zeta_{2^k}^r\zeta_8^j, \]
for some $j\in \{1,3,5,7\}$. Therefore, $f$ is gbent.

If conversely $f$ is gbent, then by Proposition \ref{HfromS} we have
\[ S_0 + S_1\zeta_{2^k} + \cdots + S_{2^{k-1}-1}\zeta_{2^k}^{2^{k-1}-1} = 2^{k-1}\mathcal{H}_f(u) = 2^{k-1}2^{\frac{n-1}{2}}\sqrt{2}\zeta_{2^k}^j, \]
for some  $0\le j\le 2^{k-1}-1$. By Lemma \ref{sqrt2}, there exists (a unique) $r$, $0\le r\le 2^{k-2}-1$, such that
\[ \sqrt{2}\zeta_{2^k}^j = \pm \zeta_{2^k}^r \pm \zeta_{2^k}^{r+2^{k-2}}. \]
Combining the two above relations, we have
\[ S_0 + S_1\zeta_{2^k} + \cdots + S_{2^{k-1}-1}\zeta_{2^k}^{2^{k-1}-1} = 2^{k-1}2^{\frac{n-1}{2}}(\pm \zeta_{2^k}^r \pm \zeta_{2^k}^{r+2^{k-2}}). \]
Therefore, $S_r = \pm 2^{k-2}2^{\frac{n+1}{2}}$, $S_{r+2^{k-2}} = \pm 2^{k-2}2^{\frac{n+1}{2}}$, and $S_t = 0$ for $t \ne r,r+2^{k-2}$, i.e.,
\[ \mathbf{S}(u)=
  \left(
    \begin{array}{c}
      S_0 \\
      \vdots \\
      S_r \\
      \vdots \\
      S_{r+2^{k-2}} \\
      \vdots \\
      S_{2^{k-1}-1} \\
    \end{array}
  \right)
  = 2^{k-2}2^{\frac{n+1}{2}}
  \left(
    \begin{array}{c}
      0 \\
      \vdots \\
      (-1)^{e_1} \\
      \vdots \\
      (-1)^{e_2} \\
      \vdots \\
      0 \\
    \end{array}
  \right),\quad e_1,e_2\in\{0,1\}. \]
By the invertibility of $H_{2^{k-1}}$, the linear system $(\ref{Sk2})$ has a unique solution for all four possibilities of $\mathbf{S}(u)$.
As now easily observed, these solutions are $(2^{\frac{n+1}{2}}H^{(r)}_{2^{k-2}},\textbf{0}_{2^{k-2}})$, $(-2^{\frac{n+1}{2}}H^{(r)}_{2^{k-2}},\textbf{0}_{2^{k-2}})$,
$(\textbf{0}_{2^{k-2}},2^{\frac{n+1}{2}} H^{(r)}_{2^{k-2}})$, and $(\textbf{0}_{2^{k-2}},-2^{\frac{n+1}{2}} H^{(r)}_{2^{k-2}})$ for
$(e_1,e_2) = (1,1), (-1,1), (1,-1)$, and $(-1,-1)$, respectively.
% \begin{remark}
% In \cite{SH2} a condition similar to that in Theorem \ref{iff1} has been shown to be sufficient for a function from $\V_n$ to $\Z_q$ to be gbent, where $n$ is even/odd and $q$
% is an arbitrary even number. However, problem posed in \cite{SH2} which regards the converse of Theorem \ref{iff1} for any $q\neq 2^k$ even still remains open.
% \end{remark}
%
%
\end{proof}

Combining Theorem \ref{iff1} and Lemma \ref{Hada}(iii) gives a characterization of the gbent property in terms of the Walsh spectral values of
the component functions. More precisely, the quadruples (four vectors) of a suitable vector space $\V_{k-1}$ which build a 2-dimensional flat
specify the component functions whose spectra satisfy certain conditions as described below. In other words, the characterization in Theorem~\ref{iff1},
which relates the spectral values of component functions to the rows of Hadamard matrices, turns out to be equivalent to a particular relation of the
Walsh spectral values for the above defined quadruples.
\begin{proposition}\label{prop1}
(i) Let $n$ be even, $k \geq3$, and represent $i=\sum_{j=0}^{k-2}i_j2^j$ for $0\le i\le 2^{k-1}-1$, with $i_j \in \{0,1\}$. Assume $g_i(x) = a_{k-1}(x) \+ i_0a_0(x)\+i_1a_1(x)\+\cdots\+i_{k-2}a_{k-2}(x)$ are bent functions, for $0\le i\le 2^{k-1}-1$. For $u\in \V_n$, the condition in Theorem~\ref{iff1}
%the relation (\ref{W=S}), which is given as
\begin{equation}
\label{Hadarow}
\mathbf{\mathcal{W}}(u) = (\mathcal{W}_{g_0}(u),\mathcal{W}_{g_1}(u),\ldots,\mathcal{W}_{g_{2^{k-1}-1}}(u)) = \pm 2^{\frac{n}{2}}H_{2^{k-1}}^{(r)}
\end{equation}
holds for some $r\in\{0,\ldots,2^{k-1}-1\}$, if and only if for any four distinct integers $j,c,l,v\in\{0,\ldots,2^{k-1}-1\}$ such that
$z_j\oplus z_c\oplus z_l\oplus z_v={\bf 0}$, the integers $\mathcal{W}_{g_j}(u),\mathcal{W}_{g_c}(u),\mathcal{W}_{{g_l}}(u),\mathcal{W}_{{g_v}}(u)\in\{-2^{\frac{n}{2}},2^{\frac{n}{2}}\}$
satisfy the equality
\begin{eqnarray}\label{wquad}
\mathcal{W}_{g_j}(u)\mathcal{W}_{g_c}(u)=\mathcal{W}_{g_l}(u)\mathcal{W}_{{g_v}}(u).
\end{eqnarray}
(ii) Similarly, when   $n$ be odd, let us assume  that  $g_i(x) = a_{k-1}(x) \+ i_0a_0(x)\+i_1a_1(x)\+\cdots\+i_{k-2}a_{k-2}(x)$ are  semi-bent functions, for any $0\le i\le 2^{k-1}-1$. Then,
\begin{equation*}
%\label{oddW=S}
\mathbf{\mathcal{W}}(u) = (\pm 2^{\frac{n+1}{2}}H^{(r)}_{2^{k-2}},\textbf{0}_{2^{k-2}})%\quad\mbox{or}\quad
%\mathbf{\mathcal{W}}(u) = \{\textbf{0}_{2^{k-2}},\pm 2^{\frac{n+1}{2}} H^{(r)}_{2^{k-2}}\}
\end{equation*}
for some $0\le r\le 2^{k-2}-1$, if and only if $\mathcal{W}_{g_j}(u) = 0$ for all $2^{k-2}\le j\le 2^{k-1}-1$ and
$\mathcal{W}_{g_j}(u) \ne  0$ for all $0 \le j\le 2^{k-2}-1$ such that for any four distinct integers $j,c,l,v\in\{0,\ldots,2^{k-2}-1\}$ with
$z_j\oplus z_c\oplus z_l\oplus z_v={\bf 0}$, the integers $\mathcal{W}_{g_j}(u),\mathcal{W}_{g_c}(u),\mathcal{W}_{{g_l}}(u),\mathcal{W}_{{g_v}}(u)\in\{-2^{\frac{n+1}{2}},2^{\frac{n+1}{2}}\}$
satisfy the equality
\begin{eqnarray}\label{wquad}
\mathcal{W}_{g_j}(u)\mathcal{W}_{g_c}(u)=\mathcal{W}_{g_l}(u)\mathcal{W}_{{g_v}}(u).
\end{eqnarray}
A similar statement is valid for $\mathbf{\mathcal{W}}(u) = (\textbf{0}_{2^{k-2}},\pm 2^{\frac{n+1}{2}} H^{(r)}_{2^{k-2}})$.
\end{proposition}

\begin{proof}
The proposition follows from Theorem \ref{iff1} and Lemma \ref{Hada}(iii).
\end{proof}

\section{Gbent conditions in terms of affine (semi-)bent spaces}
In the previous section we have provided two different characterizations of gbent property, though both are closely related to certain properties of
the component functions. The derived conditions essentially also capture the inherent properties of the affine spaces of (semi-)bent functions that
correspond to gbent functions. In this section we  specify  these affine spaces of (semi-)bent functions  and also address the affine equivalence
of gbent functions in a rigour manner.

We first develop equivalent gbent conditions in terms of affine bent spaces for even $n$. In this case, by the definition of the dual $g^*$ of a bent function $g$, the relation (\ref{wquad}) in Proposition \ref{prop1} is equivalent to
$$ (-1)^{g^*_j(u)}(-1)^{g^*_c(u)}=(-1)^{g^*_l(u)}(-1)^{g^*_v(u)},$$
for all $u\in\V_n$. Hence $g^*_j\oplus g^*_c\oplus g^*_l\oplus g^*_v=0,$ if $j,c,l,v$ satisfy $z_j\oplus z_c\oplus z_l\oplus z_v={\bf 0}$.
Observing that $g_j\oplus g_c\oplus g_l\oplus g_v=0$ if and only if $z_j\oplus z_c\oplus z_l\oplus z_v={\bf 0}$, we obtain the following
corollary from Theorem \ref{iff1} and Proposition \ref{prop1}.
\begin{corollary}
\label{dualiff}
A function $f:\V_n\rightarrow\Z_{2^k}$, $n$ even, given as $f(x) = a_0(x) + 2a_1(x) + \cdots + 2^{k-1}a_{k-1}(x)$ is gbent if and only if
\[ \mathcal{A} = a_{k-1} \oplus \langle a_0,a_1,\ldots,a_{k-2}\rangle \]
is an affine vector space of bent functions such that for any $h_0,h_1,h_2,h_3 \in \mathcal{A}$ with $h_0\+h_1\+h_2\+h_3 = 0$ we have
$h_0^*\+h_1^*\+h_2^*\+h_3^* = 0$. Equivalently, if $h_3 = h_0\+h_1\+h_2$, then $h^*_3 = h^*_0\+h^*_1\+h^*_2$.
\end{corollary}
%\textbf{What is $\langle a_0,a_1,\ldots,a_{k-2}\rangle$? We need to be define it somewhere, since the early notation of scalar product is using $\langle,\rangle$.}\\\\
%%
Corollary \ref{dualiff} generalizes an observation in \cite{m}, where the relations between octal gbent functions and a secondary construction of bent functions proposed by
Carlet \cite{c} were investigated. We state the version of this construction \cite{c} given by Mesnager in \cite{mesna}.
\begin{proposition}\cite[Th. 4]{mesna}
\label{mes}
Let $g_0,g_1,g_2,g_3$ be bent functions from $\V_n$ to $\F_2$ such that $g_0\+g_1\+g_2\+g_3 = 0$. Then the function
\[ g_0g_1 \+ g_0g_2 \+ g_1g_2 \]
is bent if and only if $g_0^*\+g_1^*\+g_2^*\+g_3^* = 0$, and its dual is $g_0^*g_1^* \+ g_0^*g_2^* \+ g_1^*g_2^*$.
\end{proposition}
%
%As we see from Theorem a gbent function $f:\V_n\rightarrow\Z_{2^k}$, $n$ even, $k\ge 3$, provides (many) quadruples of bent functions satisfying the condition in
%Proposition \ref{mes}. In fact we have
Combining Corollary \ref{dualiff} and Proposition \ref{mes} we get interesting alternative conditions for gbent functions in $\mathcal{GB}_n^{2^k}$ when $n$ is even.
\begin{corollary}
\label{evenThm}
Let $n$ be even. A function $f(x) = a_0(x) + 2a_1(x) + \cdots + 2^{k-1}a_{k-1}(x) \in\mathcal{GB}_n^{2^k}$ is a gbent function if and only if
$\mathcal{A} = a_{k-1} \+ \langle a_0,a_1,\ldots,a_{k-2}\rangle$ is an affine vector space of bent functions such that for every (pairwise distinct) $g_i,g_j,g_l\in\mathcal{A}$
the function $g_ig_j \+ g_ig_l \+ g_jg_l$ is bent.
\end{corollary}
\begin{remark}
Note that if the bent functions $g_i,g_j,g_l$ are not pairwise distinct, then $g_ig_j \+ g_ig_l \+ g_jg_l$ is trivially bent.
\end{remark}

To address the case when $n$ is odd, we show the following analog of Proposition \ref{mes} for semi-bent functions.
\begin{proposition}
\label{mesodd}
Let $g_0,g_1,g_2,g_3$ be semi-bent functions from $\V_n$ to $\F_2$ such that $g_0\+g_1\+g_2\+g_3 = 0$. Then, the function
\[ g_0g_1 \+ g_0g_2 \+ g_1g_2 \]
is semibent if and only if for all $u\in\V_n$, $\mathcal{W}_{g_i}(u) = 0$ for an even number of $i\in\{0,1,2,3\}$, and if
$\mathcal{W}_{g_i}(u) \ne 0$ for all $i\in\{0,1,2,3\}$, then
\begin{equation}
\label{wquad1}
\mathcal{W}_{g_0}(u)\mathcal{W}_{g_1}(u)=\mathcal{W}_{g_2}(u)\mathcal{W}_{{g_3}}(u),
\end{equation}
or
\begin{equation}
\label{wquad2}
|\{i\,:\,\mathcal{W}_{g_i}(u) = 2^{(n+1)/2}\}| = 1,3,\quad\mbox{but not}\quad \mathcal{W}_{g_0}(u)=\mathcal{W}_{g_1}(u)=\mathcal{W}_{g_2}(u).
\end{equation}
\end{proposition}
{\it Proof.}
By \cite[Lemma 1]{c} (see also Proposition 2 in \cite{mesna}), for (pairwise distinct) Boolean functions $g_0,g_1,g_2,g_3$ such that
$g_0\+g_1\+g_2\+g_3 = 0$, the Walsh-Hadamard transform of $g=g_0g_1 \+ g_0g_2 \+ g_1g_2$ satisfies
\[ \mathcal{W}_g(u) = \frac{1}{2}(\mathcal{W}_{g_0}(u)+\mathcal{W}_{g_1}(u)+\mathcal{W}_{g_2}(u)-\mathcal{W}_{g_3}(u)) \]
for all $u\in\V_n$. The correctness of the proposition follows then easily by checking all possible combinations of $\mathcal{W}_{g_i}(u)$,
$i\in\{0,1,2,3\}$. Note that $(\ref{wquad1})$ is equivalent to $\mathcal{W}_{g_i}(u) = -2^{(n+1)/2}$ for an even number of
$i\in\{0,1,2,3\}$. \hfill$\Box$
\begin{remark}
\label{remesodd}
If for any $u\in\V_n$ for which $\mathcal{W}_{g_i}(u) \ne 0$, $i=0,1,2,3$, the condition $(\ref{wquad1})$ always applies, then
$g = g_ig_j\oplus g_ig_l\oplus g_jg_l$ is semi-bent for any $\{i,j,l\}\subset\{0,1,2,3\}$.
If for some of  $u \in\V_n$ we have $(\ref{wquad2})$, then this is not true.
\end{remark}
%
% \begin{theorem}
% Let $n$ be odd. If $f(x) = a_0(x) + 2a_1(x) + \cdots + 2^{k-1}a_{k-1}(x) \in\mathcal{GB}_n^{2^k}$ is a gbent function, then
% $\mathcal{A} = a_{k-1} + \langle a_0,a_1,\ldots,a_{k-2}\rangle$ is an affine vector space of semibent functions such that for every (pairwise distinct)
% $g_i,g_j,g_l\in\mathcal{A}$ the function $g=g_ig_j+g_ig_l+g_jg_l$ is semibent.
% \end{theorem}

\begin{corollary}
\label{oddThm}
Let $n$ be odd. If $f(x) = a_0(x) + 2a_1(x) + \cdots + 2^{k-1}a_{k-1}(x) \in\mathcal{GB}_n^{2^k}$ is a gbent function, then
$\mathcal{A} = a_{k-1} \+ \langle a_0,a_1,\ldots,a_{k-2}\rangle =a_{k-1} \+ \mathcal{L}$ is an affine vector space of semi-bent functions such that for every (pairwise distinct)
$g_i,g_j,g_l\in\mathcal{A}$ the function $G=g_ig_j \+ g_ig_l \+ g_jg_l$ is semi-bent.
Moreover, for every $u\in\V_n$ we have
\begin{align}
\label{half=0}
\nonumber
& \mathcal{W}_g(u) = 0\;\;\mbox{if and only if}\;\; g\in a_{k-1} \+ \langle a_0,a_1,\ldots,a_{k-3}\rangle,\; \mbox{or} \\
& \mathcal{W}_g(u) \ne 0\;\;\mbox{if and only if}\;\; g\in a_{k-1} \+ \langle a_0,a_1,\ldots,a_{k-3}\rangle.
\end{align}
Conversely, if $\mathcal{A} = a_{k-1} \+ \mathcal{L}$
%\textbf{What is $\mathcal{L}$? It is clear from context, but need to say it somewhere...} {\bf Enes comment : This statement is unreadable !! What is the difference in (12), where is $a_{k-2}$ ??}
is an affine vector space of semibent functions such that for every (pairwise distinct)
$g_i,g_j,g_l\in\mathcal{A}$ the function $G=g_ig_j \+ g_ig_l \+ g_jg_l$ is semibent, and $\mathcal{A} = a_{k-1} \+ \langle a_{k-2},\mathcal{L}_1\rangle$ for some
 subspace $\mathcal{L}_1$ of $\mathcal{L}$ and $a_{k-2}\not\in\mathcal{L}_1$, with the property that for all $u \in \V_n$ we have
\begin{align}
\label{0ne0}
\nonumber
& \mathcal{W}_g(u) = 0\;\;\mbox{if and only if}\;\; g\in a_{k-1} \+ \mathcal{L}_1,\; \mbox{or}\; \\
& \mathcal{W}_g(u) \ne 0\;\;\mbox{if and only if}\;\; g\in a_{k-1} \+ \mathcal{L}_1,
\end{align}
then $f(x) = a_0(x) + 2a_1(x) + \cdots + 2^{k-3}a_{k-3}(x) + 2^{k-2}a_{k-2}(x) + 2^{k-1}a_{k-1}(x)$, $a_i\in\mathcal{L}_1$, $0\le i\le k-3$, is gbent.
\end{corollary}
\begin{proof}
Let $f(x) = a_0(x)+2a_1(x)+\cdots+2^{k-1}a_{k-1}(x)$ be a gbent function from $\V_n$ to $\Z_{2^k}$, $n$ odd, and for $0\le i\le 2^{k-1}-1$ let
$g_i(x) = i_0a_0(x) \+ i_1a_1(x) \+ \cdots \+ i_{k-2}a_{k-2}(x) \+ a_{k-1}(x)$, where $(i_0,i_1,\ldots,i_{k-2})=z_i$ is the binary representation of $i$. By Theorem \ref{iff1}(ii), $g_i$ is semi-bent for all $0\le i\le 2^{k-1}-1$ and
\begin{equation}
\label{0HH0}
\mathbf{\mathcal{W}}(u) = (\pm 2^{\frac{n+1}{2}}H^{(r)}_{2^{k-2}},\textbf{0}_{2^{k-2}})\;\;\mbox{or}\;\;
\mathbf{\mathcal{W}}(u) = (\textbf{0}_{2^{k-2}},\pm 2^{\frac{n+1}{2}} H^{(r)}_{2^{k-2}})
\end{equation}
for some $0\le r\le 2^{k-2}-1$.

Let $0 \le i < j < l < v \le 2^{k-1}-1$ be such that $z_i \+ z_j \+ z_l \+ z_v = {\bf 0}$, where $z_i = (i_0,i_1,\ldots,i_{k-2}) \in \V_{k-1}$ is
the binary representation of $i$. Since $i_{k-2}\+ j_{k-2} \+ l_{k-2} \+ v_{k-2} = 0$, the following situations can then occur:
\begin{itemize}
\item[(i)] $0 \le i,j,l,v \le 2^{k-2}-1$: In this case, either $\mathcal{W}_h(u)=0$ for all $h\in\{g_i,g_j,g_l,g_v\}$,
or $\mathcal{W}_h(u)\ne 0$ for all $h\in\{g_i,g_j,g_l,g_v\}$. In the latter case, by Proposition \ref{prop1},
$\mathcal{W}_{g_i}(u)\mathcal{W}_{g_j}(u)=\mathcal{W}_{g_l}(u)\mathcal{W}_{{g_v}}(u)$.
In both cases, by Proposition \ref{mesodd} the function $G$ is semi-bent.
\item[(ii)] $2^{k-2} \le i,j,l,v \le 2^{k-1}-1$: The same argument as for (i) applies to this case.
\item[(iii)] $0 \le i,j \le 2^{k-2}-1$, $2^{k-2} \le l,v \le 2^{k-1}-1$: In this case, exactly two of $\mathcal{W}_{g_i}(u)$,
$\mathcal{W}_{g_j}(u)$, $\mathcal{W}_{g_l}(u)$ $\mathcal{W}_{g_v}(u)$ are zero, hence by Proposition \ref{mesodd} the function
$G$ is semi-bent.
\end{itemize}
Finally, $(\ref{half=0})$  follows directly from $(\ref{0HH0})$.

To show the converse, we first note that $(\ref{0ne0})$ implies that exactly $2^{k-2}$ entries of $\mathcal{W}(u)$ are zero, all of them being located either at the first
or at the second half of $\mathcal{W}(u)$. Since we suppose that $g_ig_j \+ g_ig_l \+ g_jg_l$ is semibent for {\it all} (pairwise distinct) $g_i,g_j,g_l\in\mathcal{A}$, by Proposition \ref{mesodd} and Remark \ref{remesodd},
the nonzero half of $\mathcal{W}(u)$ equals to $\pm 2^{\frac{n+1}{2}}H_{2^{k-2}}^{(r)}$ for some $0\le r\le 2^{k-2}-1$.
As a consequence, $f$ is gbent by Theorem \ref{iff1}(ii).
\end{proof}

\subsection{Equivalence of gbent functions}
\label{chara}

We now give the complete characterization of gbent functions, both for even and odd $n$, as an algebraic object.
Similarly to the case of standard bent functions we discuss the concept of affine equivalence of gbent functions.

As already demonstrated, a gbent function $f(x) = a_0(x)+2a_1(x) + \cdots + 2^{k-2}a_{k-2}(x) + 2^{k-1}a_{k-1}(x)$ gives rise
to $\mathcal{A} = a_{k-1} \+ \langle a_0,\ldots,a_{k-2}\rangle$ which is an affine space of bent functions (semi-bent functions)
with certain properties. Thus, it is natural to  investigate its correspondence  to apparently similar class of functions,
namely to vectorial bent functions. Recall that a vectorial bent function $F:\V_n\rightarrow\V_k$, $n$ even, $k\le n/2$, is a function
\begin{equation}
\label{vecbent}
F(x) = (a_0(x),a_1(x),\ldots,a_{k-1}(x)),\quad a_i\in\mathcal{B}_n,\;0\le i\le k-1,
\end{equation}
for which every (nontrivial) component function $i_0a_0 \+ i_1a_1 \+ \cdots \+i_{k-1}a_{k-1}$, $i_j\in\{0,1\}$, $0\le j\le k-1$, is bent. Equivalently, $F$ is a
$k$-dimensional vector space of bent functions with a basis $\{a_0,a_1,\ldots,a_{k-1}\}$. Changing the basis, that is,  performing a coordinate transformation on $\V_k$ does not change the vector space. It is rather the representation in the form $(\ref{vecbent})$ that changes. In spite of a different appearance, the functions are considered to be the same. Furthermore, it is well known that  a coordinate transformation on $\V_n$ also results in a vectorial bent function, which is said to be equivalent and is not seen as a different object. For these reasons a discussion about the equivalence of gbent functions seems to be in place.

Let $f(x) = a_0(x)+2a_1(x) + \cdots + 2^{k-2}a_{k-2}(x) + 2^{k-1}a_{k-1}(x)\in \mathcal{GB}_n^{2^k}$ be a gbent function,
 $b\in\langle a_0,a_1,\ldots,a_{k-2}\rangle$, and let $B$ be an invertible $(k-1)\times(k-1)$-matrix over $\V_2$. Set
$\aa = (a_0(x),a_1(x),\ldots,a_{k-2}(x))$ and let $B\aa^T = (b_0(x),b_1(x),\ldots,b_{k-2}(x))$. Then,
\[ a_{k-1} \+ \langle a_0,a_1,\ldots,a_{k-2}\rangle \quad\mbox{and}\quad (a_{k-1} \+ b) \+ \langle b_0,b_1,\ldots,b_{k-2}\rangle \]
define the same affine space of bent functions respectively semi-bent functions. In particular, when $n$ is even,
the function
\[ f_1(x) = b_0(x)+2b_1(x)+\cdots+2^{k-2}b_{k-2}(x)+2^{k-1}(a_{k-1}(x) \+ b(x)) \]
 is also a gbent function, describing  the same object as $f$ does. One has to be little bit more careful when $n$ is odd, since then
the vector space $\mathcal{L}=\langle a_0,a_1,\ldots,a_{k-2}\rangle$ contains a  subspace $\mathcal{L}_1$ as described  in Corollary \ref{oddThm}.
Thus, for our standard representation, when $f$ is of the form $(\ref{eq:1})$, $a_{k-2}$ has to be chosen from $\mathcal{L}\setminus\mathcal{L}_1$. \\[.3em]
As for (vectorial) bent functions one can obtain seemingly new gbent functions from a given one by applying a coordinate transformation on $\V_n$.
Let $f:\V_n \rightarrow \Z_{2^k}$ and let $A$ be an invertible $n \times n$-matrix over $\F_2$. Then for $u\in\V_n$,
\begin{align*}
\mathcal{H}_{f(Ax)}(u) & = \sum_{x\in\V_n}\zeta_{2^k}^{f(Ax)}(-1)^{u\cdot x} = \sum_{x\in\V_n}\zeta_{2^k}^{f(x)}(-1)^{u\cdot A^{-1}x} \\
& = \sum_{x\in\V_n}\zeta_{2^k}^{f(x)}(-1)^{(A^{-1})^Tu\cdot x} = \mathcal{H}_{f}((A^{-1})^Tu).
\end{align*}
Hence $f(Ax)$ is gbent if and only if $f$ is gbent. Consequently, gbentness is invariant under linear coordinate transformations on $\V_n$.
From the above discussion, when $n$ is even, we may say that $f(x) = a_0(x)+2a_1(x)+\cdots+2^{k-2}a_{k-2}(x)+2^{k-1}a_{k-1}(x)$ and $f_1(x)$
%= b_0(x)+2b_1(x)+\cdots+2^{k-2}b_{k-2}(x)+2^{k-1}b_{k-1}(x)$
are equivalent if there exist $A\in GL(n,\F_2)$, $B\in GL(k-1,\F_2)$ and $b\in \langle a_0,a_1,\ldots,a_{k-2}\rangle$, such that
\[ f_1(x) = b_0(Ax)+2b_1(Ax)+\cdots+2^{k-2}b_{k-2}(Ax)+2^{k-1}b_{k-1}(Ax) \]
with $(b_0(x),b_1(x),\ldots,b_{k-2}(x)) = B\aa^T$ and $b_{k-1} = a_{k-1}\oplus b$. When $n$ is odd, we require that the coordinate transformation
induced by $B$ leaves the subspace $\mathcal{L}_1$ invariant.
%
% Consequently, with
% \[ f_1(x) = a_0(x)+2a_1(x)+\cdots+2^{k-2}a_{k-2}(x)+2^{k-1}a_{k-1}(x), \]
% also
% \[ f_2(x) = b_0(x)+2b_1(x)+\cdots+2^{k-2}b_{k-2}(x)+2^{k-1}(a_{k-1}(x)+b(x)) \]
% is gbent. (This by the way shows some observations in a recent manuscript with Stanica and Martinsen - shown there with direct calculations - in a much more elegant way.)
%

We notice that the  gbent property  does not require  require that $a_0,a_1,\ldots,a_{k-2}$ are linearly independent. Hence,  the vector space
$\mathcal{L}=\langle a_0,a_1,\ldots,a_{k-2}\rangle$ may not have ``full'' dimension $k-1$. When $n$ is even, in the extreme case $\dim(\mathcal{L})=0$,
and $f(x) = 2^{k-1}a_{k-1}(x)$ is a gbent function if  $a_{k-1}$ is a bent function. Then the image set of $f$ is two-valued taking  the values in $\{0,2^{k-1}\}$, but certainly one will not consider $a_{k-1}$ and
$2^{k-1}a_{k-1}$ as different objects. In general, it is easily verified that if
\[ f(x) = a_0(x)+2a_1(x)+\cdots+2^{k-2}a_{k-2}(x)+2^{k-1}a_{k-1}(x) \]
is a gbent function in $\mathcal{GB}_n^{2^k}$, then
\[ \tilde{f}(x) = a_0(x)+2a_1(x)+\cdots+2^{k-2}a_{k-2}(x)+2^{r-1}a_{r-1}(x) \]
is a gbent function in $\mathcal{GB}_n^{2^r}$ for any $r\ge k$.
However, this quite artificial lifted version of $f$ with a quite restricted image set, is essentially identified with
$f\in\mathcal{GB}_n^{2^k}$.

When $n$ is odd, if
\[ f(x) = a_0(x)+2a_1(x)+\cdots+2^{k-2}a_{k-2}(x)+2^{k-1}a_{k-1}(x) \]
is a gbent function in $\mathcal{GB}_n^{2^k}$, then
\[ \tilde{f}(x) = a_0(x)+2a_1(x)+\cdots+2^{k-3}a_{k-3}(x)+2^{r-2}a_{r-2}(x)+2^{r-1}a_{r-1}(x) \]
is also a gbent function in $\mathcal{GB}_n^{2^r}$, for any $r\ge k$. Again, we identify this lifted version  $\tilde{f}$ with $f$.

Let $n$ be even and suppose that the vector space $\langle a_0,a_1,\ldots,a_{r-2}\rangle$ has dimension $k-1$ for some $k\le r$.
Then there exists a matrix $B\in GL(r-1,\F_2)$ such that $B (a_0,a_1,\ldots,a_{r-2})^T = (b_0,b_1,\ldots,b_{k-2},0\ldots,0)$ for
some linearly independent $b_0,b_1,\ldots,b_{k-2}$. Hence
\[ \tilde{f}_1(x) = a_0(x)+2a_1(x)+\cdots+2^{r-1}a_{r-1}(x) \]
is equivalent to
\[ \tilde{f} = b_0(x)+2b_1(x)+\cdots+2^{k-2}b_{k-2}(x)+2^{r-1}a_{r-1}(x), \]
which is the lifted version of
\[ f = b_0(x)+2b_1(x)+\cdots+2^{k-2}b_{k-2}(x)+2^{k-1}a_{r-1}(x) \in \mathcal{GB}_n^{2^k}. \]

As a consequence of the above discussion, we can restrict ourselves to gbent functions $f(x) = a_0(x)+2a_1(x)+\cdots+2^{k-2}a_{k-2}(x)+2^{k-1}a_{k-1}(x)$
for which $a_0,a_1,\ldots,a_{k-2}$ are linearly independent. The same argument also applies to the $n$ odd case.
The following summary of our discussion is fundamental for the characterization and possibly a classification of gbent functions.
\begin{itemize}
\item For a gbent function $f(x)=a_0(x)+2a_1(x)+\cdots+2^{k-2}a_{k-2}(x)+2^{k-1}a_{k-1}(x) \in\mathcal{GB}_n^{2^k}$, the set $\{a_0,a_1,\ldots,a_{k-2}\}$ is
always linearly independent (otherwise it reduces to a gbent function in $\mathcal{GB}_n^{2^{k^\prime}}$ for some $k^\prime < k$).
\item A gbent function is independent from its representation of the form $(\ref{eq:1})$ via a basis of $\mathcal{L} = \langle a_0,a_1,\ldots,a_{k-2}\rangle$,
and the choice of the coset leader $a_{k-1}$ (for odd $n$ the existence of the distinguised subspace $\mathcal{L}_1$ of $\mathcal{L}$ has to be respected in the representation).
\end{itemize}
%
%
%
% Hence our objects of interest are $k-1$-dimensional affine spaces of bent functions
% with certain properties. \\[.5em]
% %
% (THIS IS THEN SETTLED FOR EVEN $n$, it should essentially also hold for $n$ odd, but we only have some technical description
% of these additional properties - which should also be invariant under coordinate transformations.). \\[.5em]
% %
% %
% % \begin{remark}
% % The vector space $\langle a_0,a_1,\ldots,a_{k-2}\rangle$ does not have to have ``full'' dimension $k-1$. In the extreme case it is the zero-space and the gbent
% % function is $f(x) = 2^{k-1}a_{k-1}(x)$ for a bent function $a_{k-1}$. This is the dirty trick (cheating) gbent function taking only the values $0$ and $2^{k-1}$.
% % I am pretty sure that if we do not have full dimension $k-1$, then the gbent function is also obtained from a gbent function for a smaller $k^\prime$ by lifted
% % to a function to $\Z_{2^k}$ (with some reduced value set?). (We should look at that to understand whats going on.) One should forget about these gbent functions.
% % \end{remark}
% % %
% % \begin{remark}
% % Sure one can change the basis for $\langle a_0,a_1,\ldots,a_{k-2}\rangle$ (or set of generators), and obtain many expressions for a gbent function. But they all
% % should be the ``same'' function (affine equivalent, one obtained from the other with a coordinate transformation, I think the coordinate transformation takes place
% % in $\F_2^{k-1}$, I suggest to look at that to really understand it, introduce a concept of equivalence....).
% % \end{remark}
% %
%
% %
We can now state the main theorems, the characterization of gbent functions in terms of affine (semi-)bent spaces.
\begin{theorem}
Let $n$ be even. A gbent function in $\mathcal{GB}_n^{2^k}$ is a $(k-1)$-dimensional affine vector space $\mathcal{A}$ of bent functions
such that for every $g_i,g_j,g_l\in\mathcal{A}$ the function $g_ig_j \+ g_ig_l \+ g_jg_l$ is bent.
\end{theorem}
%
%
% \begin{remark}
% I think that this $h$ is unique. Should be easy to show. NO. I can replace it with any in $L$ but not $L_1$,
% \end{remark}
%
\begin{theorem}
Let $n$ be odd. A gbent function in $\mathcal{GB}_n^{2^k}$ is a $(k-1)$-dimensional affine vector space $\mathcal{A} = a_{k-1} \+ \mathcal{L}$
of semibent functions for which $g_ig_j \+ g_ig_l \+ g_jg_l$ is semibent for every $g_i,g_j,g_l\in\mathcal{A}$, and for all $u\in\V_n$ we have
\begin{align*}
& \mathcal{W}_g(u) = 0\;\mbox{if and only if}\; g\in a_{k-1} \+ \mathcal{L}_1,\; \mbox{or} \\
& \mathcal{W}_g(u) \ne 0\;\mbox{if and only if}\; g\in a_{k-1} \+ \mathcal{L}_1,
\end{align*}
for some $(k-2)$-dimensional subspace $\mathcal{L}_1$ of $\mathcal{L}$.
\end{theorem}
%
%
% triple $(a_{k-1},a_{k-2},\mathcal{L}_1)$, where $\mathcal{L}_1$ is a $k-1$-dimensional
% vector space of Boolean functions in $\mathcal{B}_n$, $a_{k-1},a_{k-2} \in \mathcal{B}_n$, $a_{k-2}\not\in\mathcal{L}_1$ with the
% following properties: $\mathcal{A} = a_{k-1} \+ \mathcal{L} = a_{k-1} \+ \langle a_{k-2},\mathcal{L}_1\rangle$ is an affine space of
% semibent functions for which $G = g_ig_j+g_ig_l+g_jg_l$ is semibent for every (pairwise distinct) $g_i,g_j,g_l\in\mathcal{A}$ and for
% all $u\in\V_n$ we have
% \begin{align*}
% & \mathcal{W}_g(u) = 0\;\mbox{if and only if}\; g\in a_{k-1} \+ \mathcal{L}_1,\; \mbox{or} \\
% & \mathcal{W}_g(u) \ne 0\;\mbox{if and only if}\; g\in a_{k-1} \+ \mathcal{L}_1,
% \end{align*}
% \end{theorem}

\section{$\Z_q$-bent functions, vectorial bent functions and relative difference sets}
\label{RDS}

In this section, $n$ is always even, $q=2^k$.
We recall that a $\Z_q$-bent function is a function from an $n$-dimensional vector space $\V_n$ over $\V_2$ to $\Z_q$, for which
\[ \mathcal{H}_f(a,u) = \sum_{x\in\V_n}\zeta_q^{af(x)}(-1)^{u\cdot x} \]
has absolute value $2^{n/2}$ for every $u\in \V_n$ and nonzero $a\in\Z_q=\Z_{2^k}$.
Equivalently, a $\Z_q$-bent function given by its graph
$D = \{(x,f(x))\,:\,x\in\V_n\}$ is a $(2^n,2^k,2^n,2^{n-k})$-relative difference set in $\V_n\times\Z_{2^k}$.
Clearly, a $\Z_q$-bent function is always gbent. In \cite{Vect} more general vectorial $\Z_q$-bent functions
are considered. We focus on the most interesting case where the codomain is cyclic.
%
%A somewhat similar approach of naturally extending the group of characters, was
%recently used in \cite{Vect} for deriving the results on relative difference sets for vectorial gbent functions. However, the above
%extension is in our case applied to a single gbent function and the constant $a$ stands for the multiplication of the output values
%(modulo $q$) and not for the selection of the linear combinations as in the vectorial case.
%
\begin{proposition}
\label{bentcon}
A function $f(x)=a_0(x)+2a_1(x)+\cdots+2^{k-1}a_{k-1}(x)\in\mathcal{GB}_n^{2^k}$, $n$ even, is $\Z_q$-bent if and only if
$2^tf(x) = 2^ta_0(x)+2^{t+1}a_1(x)+\cdots+2^{k-1}a_{k-t-1}(x) \sim a_0(x)+2a_1(x)+\cdots+2^{k-t-1}a_{k-t-1}(x)$ is a gbent function with dimension $k-1-t$
for every $t=0,1,\ldots,k-1$.
\end{proposition}
\begin{proof}
If $f$ is $\Z_q$-bent, then $|\mathcal{H}^{(2^k)}_f(2^t,u)| = 2^{n/2}$ for every $u\in \V_n$ and $t=0,1,\ldots,k-1$ by definition.

%\textcolor[rgb]{0.98,0.00,0.00}{I deleted here that $|\mathcal{H}_f(u)| = 2^{n/2}$ is independent from from the choice of the
%primitive $2^k$th complex root of unity $\zeta_{2^k}$ used in the calculations. It is too trivial, it is only an automorphism in $\Q(\zeta)$}

For the converse, let $a = 2^tz$, for an odd integer $z$ and $0\le t\le k-1$. We have to show that $|\mathcal{H}_f(a,u)| = 2^{n/2}$ for all $u\in\V_n$.
Let $f_t(x) = a_0(x)+2a_1(x)+\cdots+2^{k-1-t}a_{k-1-t}(x)$. By assumption, for all $u\in\V_n$,
\begin{align*}
& \mathcal{H}^{(2^k)}_f(2^t,u) = \sum_{x\in\V_n}\zeta_{2^k}^{2^ta_0(x)+2^{t+1}a_1(x)+\cdots+2^{k-1}a_{k-t-1}(x)}(-1)^{u\cdot x} = \\
& \sum_{x\in\V_n}\zeta_{2^{k-t}}^{a_0(x)+2a_1(x)+\cdots+2^{k-t-1}a_{k-t-1}(x)}(-1)^{u\cdot x} = \mathcal{H}^{(2^{k-t})}_{f_t}(u),
\end{align*}
has absolute value $2^{n/2}$. Equivalently, $f_t$ is gbent, hence $\mathcal{H}^{(2^{k-t})}_{f_t}(u) = 2^{n/2}\zeta_{2^{k-t}}^r$ for some
$0\le r\le 2^{k-t}-1$ (depending on $u$). Observing that $\mathcal{H}^{(2^k)}_f(2^tz,u) = \mathcal{H}^{(2^{k-t})}_{f_t}(z,u)$ is obtained
from $\mathcal{H}^{(2^{k-t})}_{f_t}(u)$ by exchanging the primitive $2^{k-t}$-th complex root of unity $\zeta_{2^{k-t}}$ with the primitive
$2^{k-t}$-th complex root of unity $\zeta^z_{2^{k-t}}$, we conclude the proof.
\end{proof}
\begin{remark}
\label{vebefu}
As for a $\Z_q$-bent function in $\mathcal{GB}_n^{2^k}$ we require that both $f = a_0(x)+2a_1(x)+\cdots+2^{k-1}a_{k-1}(x)\in\mathcal{GB}_n^{2^k}$
and $f_1(x) = a_0(x)+2a_1(x)+\cdots+2^{k-2}a_{k-2}(x)\in\mathcal{GB}_n^{2^{k-1}}$ are gbent, thus $\langle a_0,a_1,\ldots,a_{k-1}\rangle$ is a vector
space of bent functions, i.e., a vectorial bent function.
\end{remark}

We continue with two examples of $\Z_q$-bent functions. For the first example, we employ the fact that $h(x,y) = \T_m(x\pi(y))$ is a (Maiorana-McFarland) bent function from
$\F_{2^m}\times\F_{2^m}$ to $\F_2$ if and only if $\pi$ is a permutation of $\F_{2^m}$. \\[.3em]
\begin{example} This example is based on the result in \cite[Corollary 3]{m}. Let $m$ be an integer divisible by $4$ but not by $5$, let $b,c\in\F_{2^m}^*$ with $b^4+b+1 = 0$, and let $d$ be the multiplicative inverse
of $11$ modulo $2^m-1$. Then the function $f:\F_{2^m}\times\F_{2^m}\rightarrow\Z_{2^3}$
\[ f(x,y) = \T_m(c(1+b)y^dx) + 2\T_m(c(1+b^{-1})y^dx) + 4\T_m(cy^dx) \]
is gbent \cite{m}. Now observe that $\T_m(c(1+b^{-1})y^dx)$ and $\T_m(c(1+b)y^dx) \+ \T_m(c(1+b^{-1})y^dx)$ are both Maiorana-McFarland bent functions. Hence the function
$f_1(x,y) = \T_m(c(1+b)y^dx) + 2\T_m(c(1+b^{-1})y^dx)$ is gbent in $\mathcal{GB}_{2m}^4$ by \cite[Theorem 32]{Tok}. The function $f_2(x,y) = \T_m(c(1+b)y^dx)$ is bent, thus
formally in $\mathcal{GB}_{2m}^2=\cB_{2m}$. Therefore, by Proposition \ref{bentcon}, $f(x,y)$ is $\Z_8$-bent.
\end{example}
As our second example, we analyse the $\Z_q$-bent function given in Theorem 12 in \cite{Vect} for $t=1$. The function is defined via spreads and it is not given in
the form $(\ref{eq:1})$. We start by recalling that a spread of $\V_n$, $n=2m$, is a family $S$ of $2^m+1$ subspaces $U_0,U_1,\ldots,U_{2^m}$ of $\V_n$, whose pairwise
intersection is  trivial. The classical example is the regular spread, which for $\V_n = \F_{2^m}\times\F_{2^m}$ is represented by the family
$S = \bigcup_{s\in\F_{2^m}} \{(x,sx)\,:\,x\in\F_{2^m}\} \cup \{(0,y)\,:\,y\in\F_{2^m}\}$. For the regular spread in $\V_n=\F_{2^n}$ we can take the family
$S = \{\alpha_i\F_{2^m}\,:\,i = 1,\ldots,2^m+1\}$, where $\{\alpha_i\,:\,i = 1,\ldots,2^m+1\}$ is a set of representatives of the cosets of the subgroup $\F_{2^m}^*$
of the multiplicative group $\F_{2^n}^*$ (one may take the set of the $(2^m+1)$-th roots of unity).
%\end{example}
%
\begin{example} Let $U_0,U_1,\ldots,U_{2^m}$ be the elements of a spread of $\V_n$, $n=2m$. We first construct a vectorial bent function $F$, and thereafter a $\Z_q$-bent
function $f$. We notice that  $F$ and $f$ are connected as discussed  in Remark \ref{vebefu}.

Let $\phi:\{1,2,\ldots,2^{n/2}\}\rightarrow\F_2^k$ be a balanced map, thus any $y \in \F_2^k$
has exactly $2^{n/2-k}$ preimages in the set $\{1,2,\ldots,2^{n/2}\}$. Then the function $F:\V_n\rightarrow\F_2^k$ given by
\[ F(x) = \left\{\begin{array}{l@{\quad:\quad}l}
\phi(s) & x\in U_s,\,1\le s\le 2^m,\;\mbox{and}\;x\ne 0, \\
0 & x\in U_0,
\end{array}\right.
\]
is a vectorial bent function, see e.g. Theorem 4 in \cite{cmp3}. If $a_i\in\mathcal{B}_n$, $0\le i\le k-1$, are the coordinate functions of $F$, i.e. if $F(x) = (a_0(x),a_1(x),$ $\ldots,a_{k-1}(x))$,
then $F$ is the vector space of bent functions given as $\langle a_0,a_1,\ldots,a_{k-1}\rangle$.

We now proceed with the construction of the $\Z_q$-bent function given as in \cite{Vect}. From the balanced map $\phi$, we obtain in a natural way a balanced map $\bar{\phi}$ from $\{1,2,\ldots,2^{n/2}\}$
to $\Z_{2^k}$ defined as $\bar{\phi}(s) = y_0+2y_1+\cdots+2^{k-1}y_{k-1}$ if $\phi(s) = (y_0,y_1,\ldots,y_{k-1})$. By Theorem 12 in \cite{Vect}, the function
\[ f(x) = \left\{\begin{array}{l@{\quad:\quad}l}
\bar{\phi}(s) & x\in U_s,\,1\le s\le 2^m,\;\mbox{and}\;x\ne 0, \\
0 & x\in U_0,
\end{array}\right.
\]
from $\V_n$ to $\Z_{2^k}$ is $\Z_q$-bent. Then, written in the form $(\ref{eq:1})$, $f$ is represented as $f(x) = a_0(x) + 2a_1(x) + \cdots + 2^{k-1}a_{k-1}(x)$, with the Boolean functions
$a_i$, $0\le i\le k-1$, given as above.
\end{example}

We can change the representation of the vectorial bent function $F$ by changing the basis from  $\{a_0,a_1,\ldots,a_{k-1}\}$ to $\{a_0^\prime,a_1^\prime,\ldots,a^\prime_{k-1}\}$.
The same vectorial bent function has then the representation $F(x) = \{a_0^\prime(x),a_1^\prime(x),\ldots,a^\prime_{k-1}(x)\}$. This change of the basis implies a modification of
$\phi$ and $\bar{\phi}$, and results also in an alternative formal expression for the $\Z_q$-bent function.

We emphasize that the property of being $\Z_q$-bent is much stronger than the property of being vectorial bent. $\Z_q$-bent functions are very interesting vectorial bent functions since
they correspond to two relative difference sets with parameters $(2^n,2^k,2^n,2^{n-k})$: First of all, being vectorial bent, they correspond to the relative difference set
$D=\{(x,a_0(x),a_1(x),\ldots,a_{k-1}(x))\;:\;x\in\V_n\}$ in $\V_n\times \F_2^k$, and secondly, to the relative difference set $R=\{(x,a_0(x)+2a_1(x)+\cdots+a_{k-1}(x))\;:\;x\in\V_n\}$
in $\V_n\times \Z_{2^k}$. Moreover, further relative difference sets are enclosed in such a vector space of bent functions, the relative difference sets of the bent functions of the form
$g_ig_j \+ g_ig_l \+ g_jg_l$ for some component functions $g_i,g_j,g_l$. These bent functions are in general not component functions of the vectorial bent function, hence their relative
difference sets are not projections of $D$. Here we have provided a first systematic description of this class of vectorial bent functions. There are many questions on analysis and
construction of such functions which one can investigate. We are convinced that these functions are an interesting target for future research.

\section{The dual and Gray map of gbent functions}
\label{dual}

In this section we firstly attempt to describe the dual $f^*$ of an arbitrary gbent function $f\in \mathcal{GB}^{2^k}_n$. Furthermore,
the Gray map of gbent functions is considered.

\subsection{The dual of a gbent function}

We start recalling a result of \cite{SH2}, which there is given more general for functions
in $\mathcal{GB}_n^q$, $q$ even.
%Let us recall the following result which deals with a decomposition of $\zeta^{f(x)}$ in terms of its component functions.
\begin{theorem}\cite{SH2}\label{mainth}
Let $f\in\mathcal{GB}_n^{2^k}$ be given as in $(\ref{eq:1})$, and $g_i(x) = a_{k-1}(x) \+ i_0a_0(x)\+i_1a_1(x)\+\cdots\+i_{k-2}a_{k-2}(x)$, $0\le i\le 2^{k-1}-1$, where $i=\sum_{j=0}^{k-2}i_j2^j$ and $i_j \in \{0,1\}$. Then
\begin{enumerate}
\item %$\zeta^{f(x)}$ can be represented as a linear combination of the functions $a(x)\oplus z_i\cdot A(x)$ as
\begin{eqnarray*}\label{zeta}
\zeta^{f(x)}_{2^k} = \sum^{2^{k-1}-1}_{i=0}\alpha_i (-1)^{g_i(x)},
\end{eqnarray*}
where $\alpha_i$ is given by $\alpha_i=2^{-(k-1)}H^{(i)}_{2^{k-1}}B,$ for the column matrix $B$ defined as $B=[\zeta^i_{2^k}]^{2^{k-1}-1}_{i=0}$.
\item Consequently, for any $u\in\V_n$ we have
\begin{equation*}\label{eq:Wi_lin}
\mathcal{H}_f(u)=\sum^{2^{k-1}-1}_{i=0}\alpha_i \mathcal{W}_{g_i}(u).
\end{equation*}
\end{enumerate}
\end{theorem}
% \begin{remark}
% Note that Theorem \ref{mainth} is adopted here for generalized functions given by (\ref{eq:1}) for any  $q$ even and $n$ even/odd. More generally, in \cite{SH2} is shown that
% $\mathcal{H}_f$ of any generalized function $f$ can be represented as a linear combination of $\mathcal{\mathcal{W}}_i(u)$ for any integer $q\geq 2.$
% \end{remark}

For even $n$ we will describe the dual $f^*$ of a gbent function $f\in\mathcal{GB}_n^{2^k}$ via the duals of the component functions of $f$.

\begin{theorem}\label{dualgbent}
Let $n$ be even and $f\in \mathcal{GB}^{2^k}_n$ be a gbent function given as
$$f(x)=a_0(x)+2a_1(x)+\cdots+2^{k-2}a_{k-2}(x)+2^{k-1}a_{k-1}(x),$$
for some $a_i\in \mathcal{B}_n$ $(i=0,\ldots,k-1)$, with component functions $g_j$, $0\le j\le 2^{k-1}-1$.
Then the dual $f^*\in \mathcal{GB}^{2^k}_n$ of the function $f$ is given as follows:
%\begin{itemize}
%\item[(i)] If $n$ is even, then
\begin{eqnarray}\label{dgfun1}
f^*(x)=b_0(x)+2b_1(x)+\ldots+2^{k-2}b_{k-2}(x)+2^{k-1}b_{k-1}(x),\;\;x\in \V_n,
\end{eqnarray}
where $b_{k-1}(x)=a^*_{k-1}(x)$, $b_j(x)=a^*_{k-1}(x)\oplus (a_{k-1}\oplus a_{2^j})^*(x),$ $j=0,\ldots,k-2.$
%\item[(ii)] If $n$ is odd, then
%\begin{eqnarray}\label{dgfun1}
%f^*(x)= f'(x)+(-1)^{\phi_{K_1}(x)}2^{k-3},\;\;\; x\in \V_n,
%%\left\{\begin{array}{cc}
%%                b(x)+2^{k-4}, & x\in K_0 \\
%%                b(x)-2^{k-4}, & x\in K_1
%%              \end{array}
%%\right.
%\end{eqnarray}
%where the function $f':\V_n\rightarrow \mathbb{Z}_{2^{k}}$ is given as $f'(x)=b_0(x)+2b_1(x)+\ldots+2^{k-3}b_{k-3}(x)+2^{k-1}b_{k-1}(x)$ with component functions
%$$b_{k-1}(x)=g^*_{\phi_{K_1}(x)\cdot 2^{k-2}}(x)\;\; \text{and}\;\;b_j(x)=g^*_{\phi_{K_1}(x)\cdot 2^{k-2}}(x)\oplus g^*_{\phi_{K_1}(x)\cdot 2^{k-2}+2^j}(x),$$
%where $j=0,\ldots,k-3.$
%\end{itemize}
\end{theorem}
\begin{proof}
$(i)$  From Theorem \ref{mainth} and the regularity of a gbent function $f$, we have
\[ \mathcal{H}_f(u) = \sum_{i=0}^{2^{k-1}-1}\alpha_i\mathcal{W}_{g_i}(u) = 2^\frac{n}{2}\sum_{i=0}^{2^{k-1}-1}\alpha_i(-1)^{g_i^*(u)}=2^\frac{n}{2}\zeta^{f^*(u)}_{2^{k}} . \]
Suppose that $f^*(x) = b_0(x) + 2b_1(x) + \cdots + 2^{k-1}b_{k-1}(x)$ and denote the component functions of $f^*$ by $h_i=b_{k-1}\oplus i_0 b_0\oplus \ldots i_{k-2}b_{k-2}$, $0\le i\le 2^{k-1}-1$ ($i=\sum^{k-2}_{j=0}i_j 2^{j}$). By Theorem \ref{mainth},
\[ \zeta^{f^*(x)}_{2^{k}}= \sum_{i=0}^{2^{k-1}-1}\alpha_i(-1)^{h_i(x)}. \]
Combining we get
\[ \sum_{i=0}^{2^{k-1}-1}\alpha_i(-1)^{h_i(u)} = \sum_{i=0}^{2^{k-1}-1}\alpha_i(-1)^{g_i^*(u)}. \]
Observing that $\alpha_0,\alpha_1,\ldots,\alpha_{2^{k-1}-1}$ are linearly independent $\Q(\zeta)$ (invertible matrix times $(\zeta_0,\zeta_1,$ $\ldots,\zeta_{2^{k-1}-1})$),
we obtain $h_i(x) = g_i^*(x)$, $i=0,1,\ldots,2^{k-1}-1$ (and all $x\in\V_n$). Finally, $b_{k-1}=g^*_0=a^*_{k-1}$, and with $g^*_{2^j} = b_{k-1}\+b_j$ and
$g_{2^j} = a_{k-1}\+a_j$, $j=1,\ldots,k-2$, we get
\[ b_j = a_{k-1}^* \+ (a_{k-1}\+a_j)^*,\,j=1,\ldots,k-2. \]
\end{proof}
Theorem \ref{dualgbent} generalizes the results in \cite{m} where a similar conclusion  was  stated for $k=2,3$ only.

%\subsection{The dual for odd $n$}

If $n$ is odd, then the component functions of $f$ are semi-bent, hence the description of the dual of $f$ for $n$ even
cannot transfer to $n$ odd in a straightforward manner. We leave the description of the gual od a gbent function
$f\in\mathcal{GB}_n^{2^k}$, $n$ odd, as an open problem.

\subsection{The Gray map of gbent functions}
\label{Gray}
In this section we specify the Gray image of any gbent function by showing that its Gray map is a $(k-1)$-plateaued function if $n$ is even, and $(k-2)$-plateaued function if $n$ is odd. This again generalizes the existing results that was derived in given in \cite{Gray,Tok} for $k=2,3$ and $4$.

Let $f:\V_n\rightarrow \Z_{2^k}$ be a generalized Boolean function given as
%\begin{eqnarray}\label{eq:1}
\[ f(x)=a_0(x)+2a_1(x)+2^2a_2(x)+\cdots+2^{k-1}a_{k-1}(x),\; \forall x\in \V_n. \]
%\end{eqnarray}
%
%The functions $a_i\in \mathcal{B}_n,$ $i=0,1,\ldots,k-1,$ are called the component functions of the function $f.$
The {\it generalized Gray map} $\psi(f):\mathcal{GB}^{2^k}_n\rightarrow \mathcal{B}_{n+k-1}$ of $f$ is defined by, cf. \cite{Carlet},
\begin{eqnarray}\label{greymap}
\psi(f)(x,y_0,\ldots,y_{k-2})=\bigoplus^{k-2}_{i=0}a_i(x)y_i\oplus a_{k-1}(x).
\end{eqnarray}
%\textbf{Do we need to say more on Gray maps here?}

%
%
% In this section we show that if $f\in \mathcal{GB}^{2^k}_n$ is a gbent function, then $\psi(f)$ is a $p$-plateaued function in $\mathcal{B}_{n+p}$ ($n$ even) or
% $(p-1)$-plateaued in $\mathcal{B}_{n+p}$ ($n$ odd), which consequently generalizes certain results in \cite{smgs,Grey}. First we treat the Grey image $\psi(f)$
% of the function $f.$ By adopting the notation, we start with the following result.
%
%

We start with the following result.
%\\\\
%\textbf{We can change the notation here, so instead of $p$ we can use earlier introduced integer $k$, so that we are considering here the Gray image of the %function $f$ given by (\ref{eq:1}).
%Thus, this may be polished in terms of notation, and this is more or less the copy-paste version from the very first version of article.}
%
\begin{lemma}\cite[Lemma 15]{Gray}\label{WF}
Let $n, k-1\geq 2$ be positive integers and $F:\V_n\times\V_{k-1}\rightarrow \F_2$ be defined by
$$F(x,y_0,\ldots,y_{k-2})=a_{k-1}(x)\oplus \bigoplus^{k-2}_{i=0}y_ia_i(x),\;\;\;  x\in \V_n,$$
where $a_i\in \mathcal{B}_n$, $0\leq i\leq k-1$. Denote by $A(x)$ the vectorial Boolean function $A=(a_0(x),\ldots,a_{k-2}(x))$
and let $u\in \V_n$ and $z_r\in \V_{k-1}$.
The Walsh-Hadamard transform of $F$ at point $(u,z_r)\in \V_n\times\V_{k-1}$ is then
$$\mathcal{W}_F(u,z_r)=\sum_{z_j\in \V_{k-1}}(-1)^{z_j\cdot z_r}\mathcal{W}_{a_{k-1}\oplus z_j\cdot A}(u)=H^{(r)}_{2^{k-1}}\mathcal{W}^T(u),$$
where $\mathcal{W}(u)$ is the row vector defined by (\ref{Sk2}), i.e.,
$\mathcal{W}(u)=(\mathcal{W}_0(u),\ldots,\mathcal{W}_{2^{k-1}-1}(u))$ and
$\mathcal{W}_j(u)=\mathcal{W}_{a_{k-1}\oplus z_j\cdot A}(u),$ $j=0,\ldots,2^{k-1}-1$.
\end{lemma}

We now can show that the Gray map of a gbent function in $\mathcal{GB}_n^{2^k}$ is a certain plateaued function, thus generalizing
the results on Gray maps in \cite{smgs} and \cite{Gray}  which were only given for  $q =4,8,16$.
%
%In the case when $q=2^k$, $n$ even and $p=k-1$, the following result generalizes Corollary 16 in \cite{smgs} ($q=4$),
%Theorem 16 ($q=8$) and Theorem 17 ($q=16$) in \cite{Gray}.
%
\begin{proposition}\label{greygen}
Let $f\in\mathcal{GB}_n^{2^k}$, $n$ even, be a gbent function. Then $\psi(f)$ is a $(k-1)$-plateaued function in $\mathcal{B}_{n+k-1}$, thus $\mathcal{W}_{\psi(f)} \in \{0,\pm 2^{n/2 + k-1}\}$.
\end{proposition}
\begin{proof}
By Theorem \ref{iff1}, for any $u\in \V_n$ we have $\mathcal{W}(u)=\pm 2^{\frac{n}{2}}H^{(r)}_{2^{k-1}}$ ($f$ is gbent),
for some $r\in \{0,\ldots,2^{k-1}-1\}.$ Then for arbitrary $(u,z_j)\in \V_n\times\V_{k-1}$, where
$z_j\in \V_{k-1}$, from Lemma \ref{WF} we have $F=\psi(f)$
and thus
\begin{eqnarray*}
\mathcal{W}_{\psi(f)}(u,z_j)&=&H^{(j)}_{2^{k-1}}\mathcal{W}^T(u)=H^{(j)}_{2^{k-1}}(\pm 2^{\frac{n}{2}}H^{(r)}_{2^{k-1}})^T=
\pm 2^{\frac{n}{2}}H^{(j)}_{2^{k-1}}(H^{(r)}_{2^{k-1}})^T\\
&=&\left\{\begin{array}{cc}
            \pm 2^{\frac{n}{2}+k-1}, & r=j \\
            0 & r\neq j
          \end{array}
\right.,
\end{eqnarray*}
since $H^{(j)}_{2^{k-1}}(H^{(r)}_{2^{k-1}})^T=\left\{\begin{array}{cc}
                                                                                2^{k-1}, & r=j \\
                                                                                0, & r\neq j
                                                                              \end{array}
\right.$, where $H^{(j)}_{2^{k-1}}, H^{(r)}_{2^{k-1}}$ are considered as row vectors.
%(or equivalently as vectors, due to notation from Section \ref{ifif}).
 Clearly, for $k\geq 1$ we have
$\mathcal{W}_{\psi(f)}(u,z_j)\in\{0,\pm 2^{\frac{n}{2}+k-1}\}$, which means that $\psi(f)$ is a
$(k-1)$-plateaued function in $\mathcal{B}_{n+k-1}$.
\end{proof}
\begin{proposition}\label{greygen2}
Let $f$ (defined by (\ref{eq:1})) be a gbent function in $\mathcal{GB}^{2^k}_n$, $n$ odd.
Then $\psi(f)$ is a $(k-2)$-plateaued function in $\mathcal{B}_{n+k-1}$, thus $\mathcal{W}_{\psi(f)} \in \{0,\pm 2^{\frac{n+1}{2} + k-2}\}$.
\end{proposition}
\begin{proof}
Recall that for any $u\in\V_n$ we have
$$\mathcal{W}(u)=(\pm 2^{\frac{n+1}{2}}H^{(r)}_{2^{k-2}},\textbf{0}_{2^{k-2}})\;\; \text{or}\;\;
\mathcal{W}(u)=(\textbf{0}_{2^{k-2}},\pm 2^{\frac{n+1}{2}}H^{(r)}_{2^{k-2}}),$$
for some $r\in\{0,\ldots,2^{k-2}-1\}.$ Consequently, for any $(u,z_j)\in \V_n\times \V_{k-1}$,
$$W_{\psi(f)}(u,z_j)=H^{(j)}_{2^{k-1}}\mathcal{W}^T(u)=\left\{\begin{array}{cc}
            \pm 2^{\frac{n+1}{2}+k-2}, & r\in\{j,j+2^{k-2}\} \\
            0 & r\not\in\{j,j+2^{k-2}\}
          \end{array}
\right.,$$
what completes the proof.
\end{proof}
\begin{remark}
Note that Proposition \ref{greygen} and Proposition \ref{greygen2} hold for any $q$ if $f$ is constructed by \cite[Theorem 4.1]{SH2}.
\end{remark}

\noindent
{\bf Acknowledgement.} Samir Hod\v zi\' c  is supported in part by the Slovenian Research Agency (research program P3-0384 and Young Researchers Grant). Wilfried Meidl is supported by the Austrian Science Fund (FWF) Project no. M 1767-N26. Enes Pasalic is partly supported  by the Slovenian Research Agency (research program P3-0384 and research project J1-6720).

\end{document}